\documentclass[12pt]{iopart}
\usepackage{graphicx}
\eqnobysec

\newcommand{\be}{\[}
\newcommand{\bea}{\begin{eqnarray*}}
\newcommand{\beq}{\begin{equation}}
\newcommand{\beqa}{\begin{eqnarray}}
\newcommand{\ee}{\]}
\newcommand{\eea}{\end{eqnarray*}}
\newcommand{\eeq}{\end{equation}}
\newcommand{\eeqa}{\end{eqnarray}}

\renewcommand{\d}{{\rm d}}
\def\e{{\rm e}}	

\renewcommand{\i}{{\rm i}}

\renewcommand{\P}{{\cal P}}
\renewcommand{\u}{{\bar u}}
\renewcommand{\v}{{\bar v}}
\renewcommand{\t}{{\theta}}
\renewcommand{\max}{m_{\star}}

\begin{document}

\title{Nonequilibrium phase transition in a 
non integrable zero-range process}
\author{C. Godr\`{e}che\dag\ddag}

\address{\dag\ 
Isaac Newton Institute for Mathematical Sciences,
20 Clarkson Road, Cambridge, CB3 0EH, U.K.}

\address{\ddag\ Service de Physique de l'\'Etat Condens\'e,
CEA Saclay, 91191 Gif-sur-Yvette cedex, France}

\begin{abstract}
The present work is an endeavour to determine analytically features of the stationary measure
of a non-integrable zero-range process, 
and to investigate the possible existence of phase transitions
for such a nonequilibrium model.
The rates defining the model do not satisfy the  constraints necessary for the stationary measure
to be a product measure.
Even in the absence of a drive, detailed balance with respect to this measure is violated.
Analytical and numerical investigations on the complete graph 
demonstrate the existence of a first-order phase transition
between a fluid phase and a condensed phase, where a single site has macroscopic occupation.
The transition is sudden from an imbalanced fluid where both species have densities larger than
the critical density, to a critical neutral fluid 
and an imbalanced condensate. 
\end{abstract}


\submitto{\JPA}
\maketitle	

\section{Introduction}

Zero-range processes (ZRP) are minimal models~\cite{spitz},
often used as simplified realisations of more complex processes
(for reviews, see \cite{kip,evans,lux}).
For instance they are instrumental for the understanding
of condensation transitions in driven diffusive systems~\cite{wis1,glm}.
They are closely related to urn models, 
which themselves are simplified models of 
a number of stochastic processes in Physics~\cite{urn}.
In a ZRP, particles, all equivalent, hop from sites to sites
on a lattice, with prescribed rates which only depend on the occupation of the departure site. 
The fundamental property of this process is that
the stationary measure  is explicitly known as a function of the rates,
and is a product measure~\cite{spitz,andj}.

A natural generalisation of this definition
is when two different species are allowed to coexist on each site,
again hopping with prescribed rules. However in this case 
the stationary measure is a product measure only if the rates with which 
the particles of both species leave a given site satisfy a constraint~\cite{gross,hanney} 
(see  eq.~(\ref{constraint}) below).
For short, we refer to ZRP satisfying (\ref{constraint}) as {\it integrable}.
If these rates do not satisfy the constraints,
the stationary measure is not known,
and the corresponding ZRP is a generic nonequilibrium model:
it violates detailed balance, even in the absence of a drive applied to
the system, as will be shown below.
A question of fundamental importance posed by the study of
nonequilibrium systems is the nature of their stationary state, 
and in particular the possible existence of phase transitions at stationarity.

The present work is devoted to the investigation of this question
on the particular example of a non integrable two-species ZRP.
The model arose from the study of a two-species driven diffusive
system (DDS) exhibiting,
at stationarity, condensation with coexistence between a high and a low density phase
in each individual domain~\cite{glm}.
A domain in the original DDS, i.e. a stretch of particles of the two species,
corresponds to a site in the ZRP,
while the high and low density phases correspond to the two
species of the ZRP.

We study the model on the complete graph (i.e., in the fully connected geometry), 
using analytical and numerical methods.
While for equal densities of the two species the transition
between a fluid phase and a condensed phase is continuous
(as is the case for the corresponding single-species ZRP),
for non-equal densities this transition is discontinuous.
The model exhibits a sudden phase transition from an imbalanced fluid 
where both species have densities larger than
the critical density to a neutral fluid, with densities of both species equal to the critical
density, and an imbalanced condensate. 
As a consequence reentrance is observed. The system is successively fluid, condensed, fluid,
when increasing the density of one species, holding the density of the other species fixed.
Coexistence between the two phases takes place along the transition line only. 
This study can serve as a template for the study of the one-dimensional model.

\section{Definition of the model}

\subsection{A reminder on zero-range processes}
We first give a short reminder of the definition of a ZRP.
Consider, in any dimension, a lattice of $M$ sites 
on which $N$ particles are moving.
Multiple occupancy of a site is allowed.
The dynamics consists in choosing a site at random, 
then transferring one of the particles present on this site, 
to an arrival site.
On the complete graph all sites are connected. 
The arrival site is any site chosen randomly.
In one dimension, the arrival site is one of the two nearest neighbours,
chosen with a given probability, $p$, to the right, or $q=1-p$, to the left.
The transfer of particles is done with the rate
$u_{k}$ ($k>0$),
only depending on the number $k$  of particles on
the departure site.

The fundamental property of the ZRP is that its stationary measure is known,
and is a product measure, as follows.
Let us denote by $N_i$ the random occupation of site $i$.
The stationary weight of a configuration of the system is
\beq\label{z:factor}
\P(N_{1},\dots,N_{M})=\frac{1}{Z_{M,N}}\prod_{i=1}^M p_{N_i},
\eeq
where the normalisation factor $Z_{M,N}$ reads
\beq\label{z:part}
Z_{M,N}=\sum_{N_{1}}\cdots\sum_{N_{M}}\,p_{N_{1}}\cdots p_{N_{M}}\;\delta
\left(\sum_{i}N_{i},N\right).
\eeq
For a given rate $u_k$, the factors $p_k$ obey the relation
\beq\label{rec1}
u_k\,p_k=p_{k-1}
\eeq
which leads to the explicit form 
\beq
p_0=1,\qquad p_k=\frac{1}{u_1\dots u_k}.
\label{z:pk}
\eeq

Let us emphasize two important characteristics of the ZRP (the same holding
for integrable two-species ZRP, defined below).
\begin{itemize}
\item
When the dynamics is symmetric, e.g. in the one dimensional geometry with $p=1/2$, 
or in the fully connected geometry, 
detailed balance with respect to the stationary measure
is satisfied. For the single-species ZRP the detailed balance condition reads
\be
p_k\,p_l\,u_k=p_{k-1}\,p_{l+1}\,u_{l+1},
\ee
which is precisely the property that leads to~(\ref{rec1}).\footnote{When the dynamics
is not symmetric, the condition of detailed balance is replaced by a 
condition of pairwise balance. See~\cite{lux} for a discussion of this point.}
\item
As can be seen on (\ref{z:factor}), the stationary measure is independent of the asymmetry.
As a consequence, any property of the ZRP based on the sole knowledge
of this measure is itself independent of the asymmetry.
For example, with special choices of the transfer rate $u_k$, a condensation transition 
can occur in the system. The features characterising this
phase transition are independent of the asymmetry.
\end{itemize}

This is in contrast with
the ZRP studied in the present work, where 
detailed balance is {\it not} satisfied, {\it even when the dynamics is symmetric},  
i.e., in the absence of a drive, as explained below.
In this sense this model is a generic nonequilibrium model, and the phase transition
described in the next sections is specific of a nonequilibrium system.

\subsection{The model considered in the present work}

The model considered in the present work is  
a two-species ZRP.
The general definition of a two-species ZRP is a simple extension 
of that of the usual ZRP~\cite{gross,hanney}.
Consider, in any dimension, a lattice of $M$ sites 
with $n$ particles of type 1, $m$ particles of type 2.
The dynamics consists in choosing a site at random, 
then transferring one of the particles present on this site, 
of one of the species chosen at random, to an arrival site.
The transfer of particles is done with rates
 $u_{k,l}$ ($k>0$) for a particle of the first species, 
and $v_{k,l}$ ($l>0$) for a particle of the other species,
where $k$ and $l$ are respectively the number of particles of each species
on the departure site.

At variance with the case of single-species ZRP where the stationary measure is
a product measure for any choice of the transfer rate, 
for a two-species ZRP this property holds only if the following constraint 
on the rates $u_{k,l}$ and $v_{k,l}$
is satisfied~\cite{gross,hanney}:
\beq\label{constraint}
u_{k,l}\,v_{k-1,l}=v_{k,l}\,u_{k,l-1}.
\eeq 
In the present work we choose rates which violates this constraint.
As a consequence, nothing a priori is known  on the nature of the stationary measure
of the model.
The rates read
\beq\label{rates}
u_{k,l}=1+\frac{b}{l},\qquad
v_{k,l}=1+\frac{b}{k},
\eeq
where $b$ is a given parameter, which plays the role of inverse temperature~\cite{lux}.
We also set
$u_{k,0}=v_{0,l}=1+b$ in order to complete the definition of the process.
These rates favour equality of the number of particles of both species
on each site.
This choice aims at reproducing a feature of the original DDS, where
inside the domains coexistence between a high and a low density phase takes place.
The model was first introduced in~\cite{glm}, and studied in the equal density case.
This is reviewed and extended below. We then focus on the non-equal density case.

\section{The case of two sites}

We begin by considering the case where the system is made of two sites.
This case shares many common features with the complete model and serves
as a useful preparation for the rest.

A configuration of the system is entirely specified by the numbers $k$ and $l$
of particles of each species on site 1,
since the number of particles on site 2 are then just equal to
$n-k$ and $m-l$.
Therefore the  weight of a configuration of the system is given by
the probability $f_{k,l}(t)$ that site 1 contains
$k$ particles of one species, and $l$ particles of the other species, at time $t$.
It obeys the master equation
\begin{eqnarray}\label{master2}
\frac{\d f_{k,l}(t)}{\d t} 
&=&
u_{k+1,l}\,f_{k+1,l}(1-\delta_{k,n})+v_{k,l+1}\,f_{k,l+1}(1-\delta_{l,m})\nonumber\\
&+&u_{n-k+1,m-l}\,f_{k-1,l}(1-\delta_{k,0})
+v_{n-k,m-l+1}\,f_{k,l-1}(1-\delta_{l,0})\nonumber\\
&-&\left[u_{k,l}+v_{k,l}
+u_{n-k,m-l}
+v_{n-k,m-l}\right]f_{k,l},
\end{eqnarray}
where it is understood that $u_{0,l}=v_{k,0}=0$.
This is the master equation of a biased random walk in the
rectangle $0\le k\le n$, $0\le l\le m$, with reflecting boundary conditions.

We would like to know the stationary solution
$f_{k,l}$ of this equation, for {\it any choice of rates $u_{k,l}$, $v_{k,l}$}.
It turns out that this question is already too hard to answer for the seemingly simple
problem of a two-site model.
We must content ourselves of the knowledge of the stationary solution
for the class of processes fulfilling the constraint~(\ref{constraint}).
Indeed, proviso the rates fulfill this constraint, the stationary distribution is given by
\beq\label{factor}
f_{k,l}=\frac{p_{k,l}\,p_{n-k,m-l}}{Z_{M,n,m}},
\eeq
where 
\beqa\label{rec}
p_{k,l}\,u_{k,l}=p_{k-1,l}\nonumber \\
p_{k,l}\,v_{k,l}=p_{k,l-1},
\eeqa
and 
$Z_{M,n,m}$ is a normalisation (the partition function).
Relations~(\ref{rec})  can be iterated, thus
determining the $p_{k,l}$ in terms of the rates.
Eqs.~(\ref{factor}) and (\ref{rec}) generalize eqs.~(\ref{z:factor}) and (\ref{rec1}).

The method used in~\cite{gross,hanney} to obtain these results consists
in making the ansatz~(\ref{factor}), carry this form into the master equation,
which leads to~(\ref{rec}), which itself imposes~(\ref{bis}) as a compatibility
relation.

We wish to bring an independent and complementary  viewpoint to this issue.
We first note that the dynamics between the two sites
is  symmetric.
We therefore question the possibility for the process to be reversible
in time, and the consequences thereby.
Reversibility is equivalently the property that the process obeys
detailed balance with respect to the stationary measure, 
or otherwise stated that the system is at equilibrium.
We proceed as follows.\\
(i)
We first determine the stationary distribution $f_{k,l}$
when 
detailed balance is obeyed.
Consider the transitions from $\{k,l\}$
to $\{k+1,l\}$ and back, and 
 from $\{k,l\}$
to $\{k,l+1\}$ and back.
Detailed balance requires
\bea
u_{n-k,m-l}\,f_{k,l}=u_{k+1,l}\,f_{k+1,l},\\
v_{n-k,m-l}\,f_{k,l}=v_{k,l+1}\,f_{k,l+1}.
\eea
It is readily found that a solution of these equations is given by (\ref{factor})
and (\ref{rec}).\\
(ii) 
We now determine,  by yet another path, the conditions on the rates 
for the model to satisfy reversibility.
We use the Kolmogorov criterion~\cite{kol,kel} which is a 
necessary and sufficient condition 
for a Markov process to be reversible.
This condition states that the product of the transition rates along
any cycle in the space of configurations
should be equal to the product of the transition rates for the 
reverse cycle.
In the present case, the space of configurations is the rectangle $0\le k\le n$,
$0\le l\le m$.
Taking the cycle 
\be
(k,l)\to(k,l-1)\to(k+1,l-1)\to(k+1,l)\to(k,l),
\ee
then the cycle in reverse order,
the Kolmogorov condition leads to the equation
\beq\label{kol}
\frac{u_{k+1,l}\,v_{k,l}}{u_{k+1,l-1}\,v_{k+1,l}} =
\frac{u_{n-k,m-l}\,v_{n-k,m-(l-1)}}
{u_{n-k,m-(l-1)}\,v_{n-(k+1),m-(l-1)}}.
\eeq
The two sides of this equation should be satisfied independently.
This imposes that
\beq\label{bis}
u_{k,l}\,v_{k-1,l}=u_{k,l-1}\,v_{k,l},
\eeq
which is the constraint (\ref{constraint}).%
\footnote{Eq.~(\ref{kol}) can be satisfied by imposing
symmetry relations on the rates:
\be
u_{k+1,l}=u_{n-k,m-l},\nonumber\\
v_{k,l+1}=v_{n-k,m-l}.
\ee
The corresponding stationary measure is uniform:
\be
f_{k,l}=\frac{1}{(n+1)(m+1)},
\ee
and detailed balance is obeyed.
We discard this solution because the rates would then also depend on the arrival site.}

To summarize, reversibility implies stationary product measure, eqs.~(\ref{factor}),
(\ref{rec}), and a constraint on the rates, eq.~(\ref{bis}).
The reciprocal statement holds.  
The proof follows easily from the fact that a Markov process with a finite configuration
space has a unique stationary solution. 
We leave it to the reader.

The physical interpretation of the results above is that
when the system is at equilibrium, its energy
is equal to the sum of the energies of two independent sites.

Conversely, for a choice of rates violating (\ref{bis}),
as is the case for the model studied here, the model is 
not reversible, the stationary measure does not take
the simple form (\ref{factor}), and is not known a priori.
In other words, for a general choice of rates, the two-site
model can have an arbitrarily complex
stationary measure. In this sense it represents an example of 
a minimal nonequilibrium system.

\section{The model on the complete graph}

The virtue of considering the fully connected geometry,
in the thermodynamical limit of an infinite system,
is that it leads to analytical 
results on the model.
The rest of the paper is devoted to this case.

Consider again the single-site occupation probability $f_{k,l}(t)$, that is
the probability that a generic site contains
$k$ particles of one species, and $l$ particles of the other species, at time $t$.
Conservation of probability and particle numbers imposes 
$\sum_{k,l}f_{k,l}(t)=1$
and
\beq
\sum_{k=1}^{\infty }k\,f_{k}(t) =\rho_1, \qquad
\sum_{l=1}^{\infty }l\,f_{l}(t) =\rho_2 ,
\label{rho} 
\eeq
where for a large system, densities are defined as $\rho_1=n/M$, $\rho_2=m/M$,
and where the marginals are denoted by
$f_k=\sum_{l }f_{k,l}$, and
$f_l=\sum_{k }f_{k,l} $.

The master equation for the temporal evolution of
$f_{k,l}(t)$ reads
\begin{eqnarray}\label{recgene}
\frac{\d f_{k,l}(t)}{\d t} 
&=&
u_{k+1,l}\,f_{k+1,l}+v_{k,l+1}\,f_{k,l+1}\nonumber
\\&+&\u_t\,f_{k-1,l}(1-\delta_{k,0})+\v_t\,f_{k,l-1}(1-\delta_{l,0})\\
&-&\left[u_{k,l}+v_{k,l}+\u_t+\v_t\right]f_{k,l}\;,
\nonumber
\end{eqnarray}
where
\be
\u_t =\sum_{k,l}u_{k,l}\,f_{k,l}, \qquad
\v_t =\sum_{k,l}v_{k,l}\,f_{k,l}
\ee
are the mean rates at which a particle arrives on a site
($k\rightarrow k+1 $) or ($l\rightarrow l+1 $).
Equation~(\ref{recgene}) is the master equation for a biased random walk in the quadrant
$k, l\ge0 $, with reflecting boundary conditions on the axes.

We wish to determine the stationary solution of this equation.
We follow the same line of thought as in the previous section.
We show that the stationary distribution $f_{k,l}$ has a known
closed-form expression only if reversibility is assumed.
Indeed, using the detailed balance conditions
\bea
f_{k+1,l}\,u_{k+1,l}=\u\,f_{k,l},\\
f_{k,l+1}\,v_{k,l+1}=\v\,f_{k,l},
\eea
it is easy to derive the following explicit expression for the stationary 
distribution $f_{k,l}$:
\beq\label{fklmf}
f_{k,l}=\frac{p_{k,l}\,\u^k\,\v^l}{\sum_{k,l} p_{k,l}\,\u^k\,\v^l},
\eeq
where the $p_{k,l}$ are given by (\ref{rec}), and $\u$ and $\v$ are the
stationary mean hopping rate. 

Let us also show that, as for the two-site system, the constraint~(\ref{bis})
is a consequence of imposing reversibility.
Indeed, the space of configurations is the quadrant $k,l\ge0$.
Taking the cycle 
\be
(k,l)\to(k,l-1)\to(k+1,l-1)\to(k+1,l)\to(k,l),
\ee
then the cycle in reverse order,
the Kolmogorov condition implies
\be
v_{k,l}\,\bar u\, \bar v\, u_{k+1,l}=\bar u\, v_{k+1,l}\,u_{k+1,l-1}\,\bar v,
\ee
which yields~(\ref{bis}).

As for the two-site system, we conclude that conversely, when (\ref{bis}) is violated,
as is the case with the choice of rates~(\ref{rates}),
the stationary distribution remains unknown.
The present work is an endeavour to determine features of the stationary measure
of the model for a thermodynamical system,
and to investigate the possible existence of nonequilibrium phase transitions.

In the case of an integrable
two-species ZRP, the fugacities are functions of the densities (see eq.~(\ref{fklmf})).
The duality fugacity-density is replaced here by the duality mean hopping rate-density.


\section{Criticality}
As will appear clearly as we proceed, the critical point for this model is unique, and
corresponds to taking $\u=\v=1$.

\subsection{Continuum limit: universal properties}
Let us first consider the
continuum limit of the stationary equation,
in the asymptotic regime where $k$ and $l$ are large, and setting
$\u=\e^\mu$, $\v=\e^\nu$, where $\mu$ and $\nu$ are small.
Expanding $f_{k,l}$ to second order, we obtain
\beq
\frac{\partial^2 f_{k,l}}{\partial k^2}+\frac{\partial^2 f_{k,l}}{\partial l^2}
+\frac{\partial f_{k,l}}{\partial k}\left(\frac{b}{l}-\mu\right)
+\frac{\partial f_{k,l}}{\partial l}\left(\frac{b}{k}-\nu\right)=0.
\label{limcont}
\eeq

\begin{figure}[htb]
\begin{center}
\includegraphics[angle=-90,width=.7\linewidth]{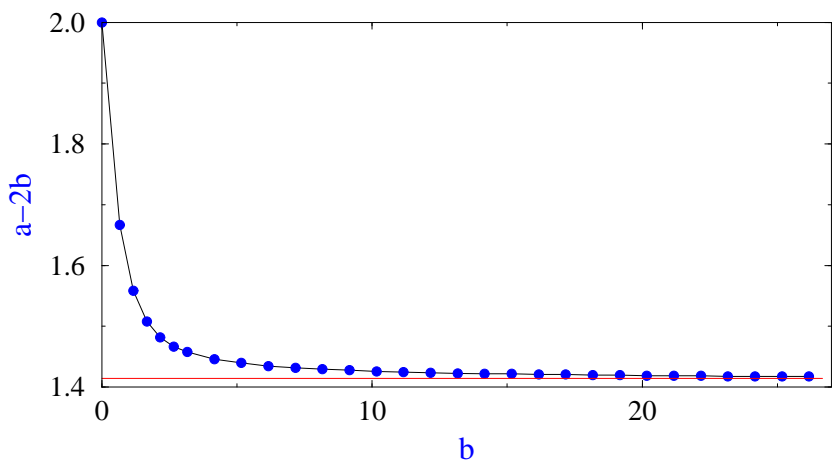}
\caption{\small
Decay exponent $a$ as a function of $b$. 
At large values of $b$, $a\approx2 b+\sqrt{2}$. 
}
\label{f1}
\end{center}
\end{figure}

At criticality, i.e. when $\mu=\nu=0$,  eq.~(\ref{limcont})
becomes scale invariant, and reads
\begin{equation}\label{crit}
\frac{\partial^2 f_{k,l}}{\partial k^2}+\frac{\partial^2 f_{k,l}}{\partial l^2}
+b\left(\frac{1}{l}\frac{\partial f_{k,l}}{\partial k}+
\frac{1}{k}\frac{\partial f_{k,l}}{\partial l}\right)
=0.
\end{equation}
Using polar coordinates: $k=r \cos\theta$, 
$l=r \sin\theta$, with $0\le \t\le\pi/2$, 
this equation is transformed into
\begin{equation}
\frac{\partial^2 f(r,\theta)}{\partial r^2}
+\frac{1}{r^2}\frac{\partial^2 f(r,\theta)}{\partial \theta^2}
+\frac{1}{r}\frac{\partial f(r,\theta)}{\partial r}
\left( 1+\frac{2b}{\sin 2\theta}\right)
=0.
\label{critpol}
\end{equation}
Now, setting $f(r,\theta)=r^{-a}g(\theta)$, we find an equation for the
angular function $g(\theta)$:
\beq\label{eq_g}
\frac{\d^2 g(\theta)}{\d \theta^2}
+a\left( a-\frac{2b}{\sin 2\theta}\right)g(\theta)
=0.
\eeq
The unknown decay exponent $a$ is determined by the boundary conditions imposed on
$g(\t)$, which are the quantisation conditions 
of this Schr\"odinger equation.
Indeed, $g(\t)$ is positive for $0\le\t\le\pi/2$, symmetric with respect to $\pi/4$
and must vanish for $\t=0$ or $\t=\pi/2$.
For special values of $b$, exact solutions of eqs.~(\ref{critpol}) 
can be found:
\beqa\label{beq0}
f(r,\t)=& r^{-2}\sin\t\cos\t\quad&(b=0),\\\label{beq1}
f(r,\t)=& r^{-3}\sin\t\cos\t(\sin\t+\cos\t)\quad&(b=2/3).
\eeqa
For $b=0$ the original model has no critical behaviour, hence formally $a=0$.
On the other hand the prediction of the continuum limit for the decay exponent,
in the limit $b\to0$,  is $a=2$.
The decay exponent $a$ is discontinuous at $b=0$.

For a generic value of $b$, the decay exponent $a$ is determined by
numerical integration of the differential equation~(\ref{eq_g}).
At large values of $b$, the behaviour of this exponent can be obtained analytically.
Indeed, expanding the potential term in (\ref{eq_g}) to second order
around  its minimum, located at $\pi/4$, yields the equation of a harmonic oscillator,
with coupling constant $\omega=2\sqrt{a b}$ and energy $a(a-2 b)/2$:
\be
g''(x)+a(a-2b-4b x^2)g(x)=0
\ee
where $x=\pi/4-\t$.
Imposing that the ground state energy be equal to $\omega/2$ yields
the asymptotic quantisation condition $a=2 b+\sqrt{2}$. (See figure~\ref{f1}.)
 
As a consequence, we find the remarkable result that,
at criticality,
the marginal distributions $f_k$ and $f_l$ 
decay as power laws at large occupations, with a non-trivial exponent equal to $a-1$.
The same holds for $p_m$, with $m=k+l$, the distribution of the total 
number of particles on a site, 
\be
p_m=\sum_{k=0}^m f_{k,m-k}\sim m^{-(a-1)}.
\ee

Note finally that both the function $g(\t)$ and the exponent $a$ are universal,
and only depend on $b$.


\subsection{Discrete equations: critical density}
\label{discrete}

The determination of the non universal critical density $\rho_c$ of both species,
where 
\be
 \rho_c=\sum_{k=1}^{\infty }k\,f_{k}=
\sum_{l=1}^{\infty }l\,f_{l} ,\qquad (\u=\v=1),
\ee
requires the knowledge of the stationary solution $f_{k,l}$ of the discrete eqs.~(\ref{recgene}).
These are integrated numerically  with $\u=\v=1$, 
using the following method.
We truncate these equations
at a given value of $k+l$ denoted by $\max$, which plays the role of a cut-off.
We solve the linear system $A\,F=I$, where $F$ is the column matrix of the occupation
probabilities $f_{k,l}$, $I$ is the matrix containing the inhomogeneous term $f_{0,0}$,
itself determined at the end of the computation by normalisation, and
$A$ is the matrix deduced from the stationary equations.
We impose the boundary conditions $f_{k,l}=0$ outside the triangle delimited by 
$k=0$, $l=0$, and $k+l=m_\star$.
The maximal value of the cut-off $\max$ attainable is limited by the size of the matrices involved. 
For example, taking $\max=160$ corresponds to a linear system of order $13040$.

\begin{figure}[htb]
\begin{center}
\includegraphics[angle=-90,width=.7\linewidth]{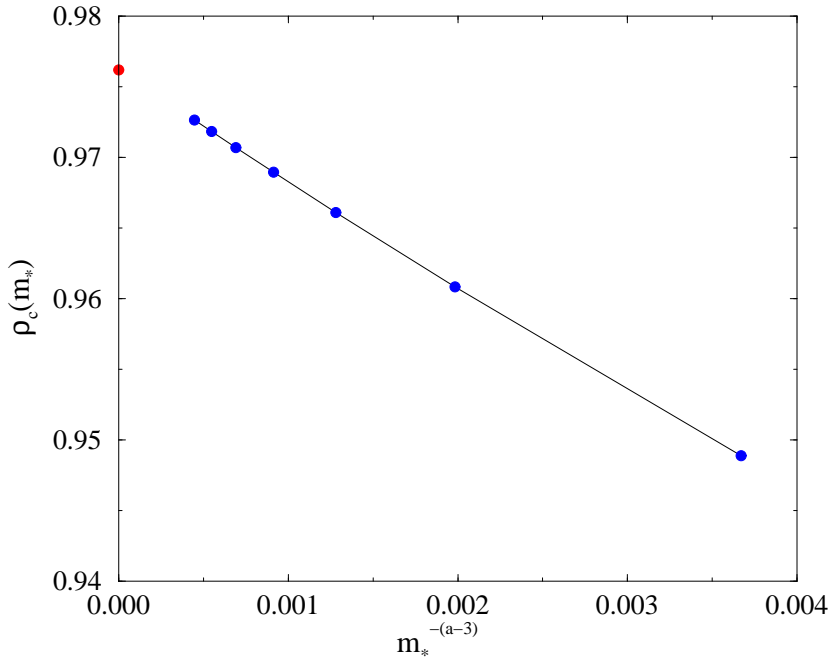}
\caption{\small
Determination of the critical density by extrapolation of the  data
for $m_\star=40,60,\ldots,160$. 
The circle on the vertical axis is the extrapolated value for $\rho_c$.
($b=3/2$, $a=4.520\ldots$,
$\u=\v=1$.)
}
\label{f4}
\end{center}
\end{figure}

As an illustration we take $b=3/2$, 
corresponding to a value of the decay exponent $a\approx4.520$.
Extrapolating the data for several values
of $m_\star$, 
using the estimate
\be
\rho_c-\rho_c({\max})\approx\int_{\max}^\infty \d m\, m\, p_m
\sim\max^{-(a-3)},
\ee
as depicted in figure~\ref{f4}, leads to $\rho_c\approx0.976$.

The theoretical prediction for 
the critical decay exponent of $p_{m}$, or $f_k\equiv f_l$, 
agrees perfectly well with numerical measurements.

\section{Fluid phase}

For non zero values of $\mu$ and $\nu$ the system is driven away from criticality.
We begin by investigating exponentially decreasing solutions of the continuum limit 
stationary equation~(\ref{limcont}).
We then study those of the discrete stationary equations.
We finally determine the region of existence of such solutions.

\subsection{Stationary solutions at exponential order: continuum limit}

If we content ourselves of the knowledge of the stationary solutions
at exponential order, that is retaining their exponential dependence only,
and discarding any prefactor, then terms containing $b$ can be neglected. 
Equation~(\ref{limcont}) now reads
\beq
\frac{\partial^2 f_{k,l}}{\partial k^2}+\frac{\partial^2 f_{k,l}}{\partial l^2}
-\mu\frac{\partial f_{k,l}}{\partial k}
-\nu\frac{\partial f_{k,l}}{\partial l}
=0.
\label{massive}
\eeq
Setting
$f_{k,l}=\e^{\frac{1}{2}(\mu\,k+\nu\,l)}\,h_{k,l}$
in (\ref{massive}) yields
\be
\frac{\partial^2 h_{k,l}}{\partial k^2}+\frac{\partial^2 h_{k,l}}{\partial l^2}
-\frac{\mu^2+\nu^2}{4} h_{k,l}
=0.
\ee
Changing to polar coordinates
and setting $h(r,\t)=u(r)v(\t)$, we obtain, after rescaling $r$ by $\sqrt{\mu^2+\nu^2}/2$,
\beq
r^2 u''(r)+r u'(r)-(r^2+n^2)u(r)=0,
\eeq
where $n$ is to be determined, and
\be
v''(\t)+n^2v(\t)=0.
\ee
Imposing $v(\t)=0$ for $\t=0$ and $\pi/2$ leads to $v(\t)=\sin 2\t$, hence 
$n=2$.
The solution of the differential equation for $u(r)$ is the Bessel function
\be
u(r)=K_2\left(\sqrt{\mu^2+\nu^2}\ \frac{r}{2}\right).
\ee
Finally, the solution, when $b\to0$, reads, up to a normalising constant,
\beq\label{b0}
f(r,\t)=
K_2\left(\sqrt{\mu^2+\nu^2}\ \frac{r}{2}\right)
\e^{\frac{r}{2}(\mu\cos\t+\nu\sin\t)}\,\sin 2\t.
\eeq
This solution encompasses all three regimes where
$\mu r\sim 1$, $\mu r\ll1$, or $\mu r\gg1$.
Simplified expressions are obtained 
in the two latter cases:

$\bullet$ For small values of the argument, we have $K_2(x)\sim 1/x^2$.
We thus find
\be
f(r,\t)\approx {\rm const.}\ r^{-2} \sin 2\t, \qquad (\mu r\ll 1),
\ee
which matches consistently with the critical solution~(\ref{beq0}) for $b\to0$.

$\bullet$ For large values of the argument, we have
$K_2(x)\approx \sqrt{\pi/2x}\,\e^{-x}$, hence we obtain the
asymptotic behaviour
\beq
f(r,\t)\approx {\rm const.}\ r^{-1/2}\,\e^{-r P(\t)}\sin 2\t, \qquad (\mu r\gg 1),
\label{asympt}
\eeq
where 
\beq
P(\t)=\frac{1}{2}\left(\sqrt{\mu^2+\nu^2}-\mu\cos\t-\nu\sin\t\right).
\label{ptheta}
\eeq
For any values of $\mu$ and $\nu$ non simultaneously positive, $P(\t) $ is positive,
$f$ is exponentially decaying, corresponding to a fluid phase.
When $\mu$ and $\nu$ are simultaneously positive, 
$P(\t)$ vanishes at an angle $\t$ satisfying $\tan\t=\nu/\mu$.
For such a value of $\t$, the function $f(r,\t)\sim r^{-1/2}$ is not normalisable.
The whole region $\mu>0$ and $\nu>0$ is therefore 
non physical. 

%

\subsection{Stationary solutions at exponential order: discrete equations }
\label{elim}

In order to investigate exponentially decaying solutions
beyond the continuum limit, we consider again the discrete stationary equations.
As above, for $k$ and $l$ large, we can neglect terms containing $b$, thus
obtaining
\beq
f_{k+1,l}+f_{k,l+1}
+\u\,f_{k-1,l}+\v\,f_{k,l-1}
-(2+\u+\v)f_{k,l}=0.
\label{stat}
\eeq
Introducing the generating function $\hat f(x,y)=\sum f_{k,l}x^ky^l$,
we get from (\ref{stat}) 
\beq\label{eq:fhat}
D(x,y)\,\hat f(x,y)
=
A(x,y),
\eeq
where 
\beq\label{eq:D}
D(x,y)=x^{-1}+y^{-1}-2+\u(x-1)+\v(y-1).
\eeq
The locus of singularities of $\hat f$ is thus given by $D(x,y)=0$.
The right-hand side, $A(x,y)$, comes from the contribution of the boundary terms
$\hat f(0,y)$ and $\hat f(x,0)$
\footnote{For the generic case $b>0$, $A(x,y)$ is finite on $D(x,y)$,
while if $b=0$, the singularities of $A(x,y)$ cancels those of $D(x,y)$
in such a way that the resulting expression for $\hat f(x,y)$ has just a simple 
pole in both complex variables $x$ and $y$:
\be
\hat f(x,y)=\frac{(1-\u)(1-\v)}{(1-\u x)(1-\v y)},\qquad (b=0).
\ee
}.
By inversion we have
\be
f_{k,l}=\oint\frac{\d x}{2\i\pi x}\frac{\d y}{2\i\pi y}\hat f(x,y)\,x^{-k}y^{-l}.
\ee
At large $k$ and $l$, $f_{k,l}$ can be estimated by taking the saddle point of
this expression, yielding
\be
f_{k,l}\sim x^{-k}y^{-l}\equiv \e^{-rP(\t)},
\ee
where
$x,y$ is a point on the curve $D(x,y)=0$, and $P(\t)=\cos \t\ln x+\sin \t \ln y$.
Extremising $P(\t)$ on this curve with respect to the variables $x$ and $y$,
i.e. the expression $P(\t)-\lambda D(x,y)$, where $\lambda$
is a Lagrange multiplier, leads, together with $D(x,y)=0$, to three equations, 
which determine $\lambda$, $x$ and $y$, hence $P(\t)$.

Let us first check that this method leads to the expected result~(\ref{ptheta}) 
in the particular case of the
continuum limit.
Set $x=\e^s$, $y=\e^t$, $\u=\e^\mu$, $\v=\e^\nu$, 
with $\mu$ and $\nu$ small.
The equation for $D(x,y)$ reads: 
\be
s^2+t^2+\mu\,s+\nu\,t=0.
\ee
Extremising $P(\t)-\lambda D(x,y)$ with respect to $s$ and $t$ yields
\bea
\lambda\cos \t-2s+\mu&=&0,\\
\lambda\sin \t-2t+\nu&=&0.
\eea
The three former equations lead, after some algebra, to eq.~(\ref{ptheta}).

\begin{figure}[htb]
\begin{center}
\includegraphics[angle=-90,width=.4\linewidth]{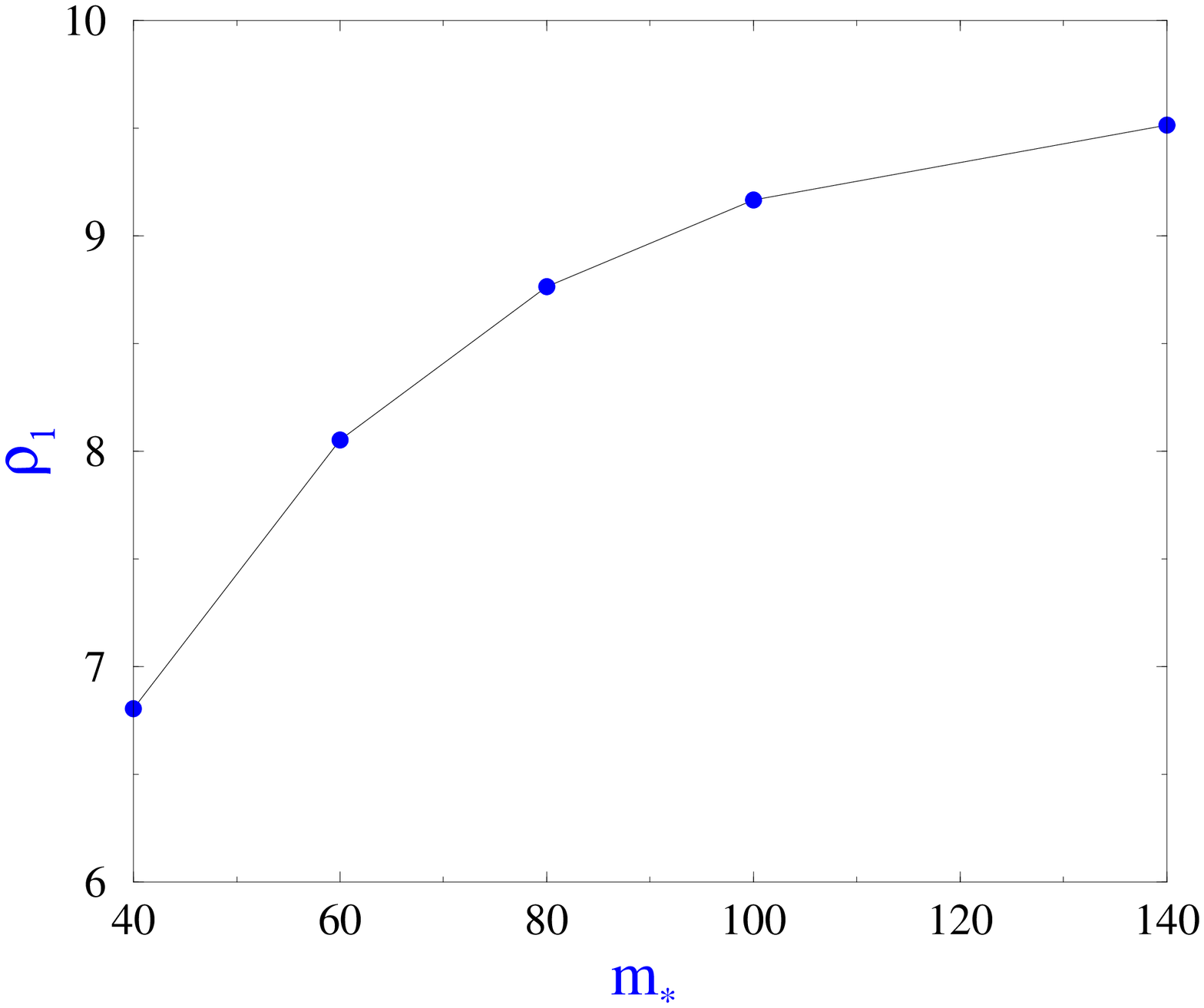}
\includegraphics[angle=-90,width=.4\linewidth]{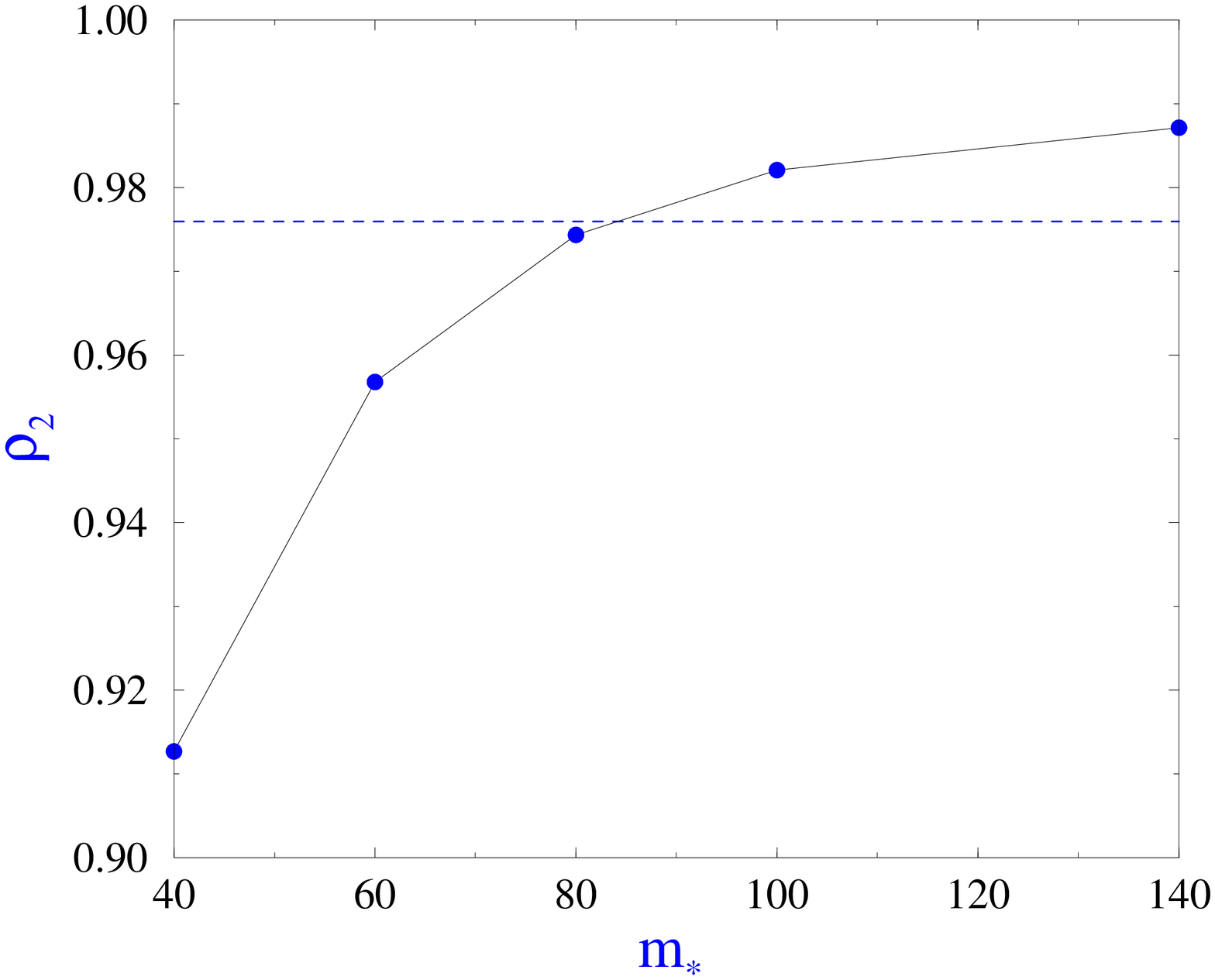}
\caption{\small
Densities as functions of $m_\star$ ($\u=1.89$, $\v=0.66$, $b=3/2$). 
The dashed line corresponds to $\rho_c$.
}
\label{fa1}
\end{center}
\end{figure}

\begin{figure}[htb]
\begin{center}
\includegraphics[angle=-90,width=.7\linewidth]{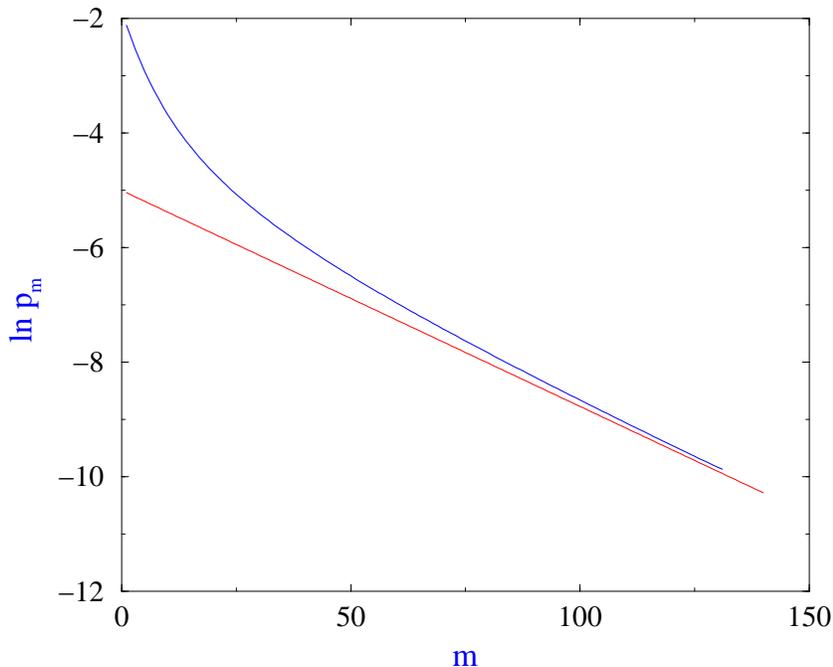}
\caption{\small
Distribution of the total single-site occupation $p_m$ as a function of $m$ ($m\le\max=140$).
Line: theoretical prediction for the coefficient of the exponential decay.
($\u=1.89$, $\v=0.66$, $b=3/2$.)
}
\label{fa2}
\end{center}
\end{figure}

The general case leads to lengthy expressions for $P(\t)$.
We give the results of this method for a particular example.
We choose $\u=1.89$, $\v=0.66$ in the stationary equations, which
we integrate by solving the linear system, as explained in section~\ref{discrete}, 
for increasing values of the cut-off $m_\star$.
The resulting values of the densities $\rho_1$ and $\rho_2$ are plotted in 
figure~\ref{fa1}.
Clearly  $\rho_2$ is larger than $\rho_c$\footnote{
The values of $\u$ and $\v$ were precisely chosen to serve this purpose.
We first integrated the master equation~(\ref{recgene}) numerically for 
$b=3/2$, corresponding to $\rho_c\approx0.976$, and for
values of the densities $\rho_1=10$ and $\rho_2=1$.
The stationarity values $\u\approx1.894$, $\v\approx0.661$ were thus obtained.
}. 
Figure~\ref{fa2} depicts $\ln p_m$, as obtained
by the same method, together with the theoretical
prediction for the coefficient of the exponential decay:
\be
p_m=\int r\,\d r\d \theta\,\e^{-rP(\t)}\delta\left ( r-\frac{m}{\cos \t+\sin\t}\right)
\sim\e^{-m \frac{P(\t_0)}{\cos \t_0+\sin\t_0}}
\ee
where $\t_0$ denotes the value of the angle such that the argument of the
exponential is minimum.
In the present case, $\t_0=0$, and $p_m\sim\e^{-0.0372\, m}$.

\subsection{Domain of existence of the fluid phase in the $\u-\v$ plane}

\begin{figure}[htb]
\begin{center}
\includegraphics[angle=-90,width=.7\linewidth]{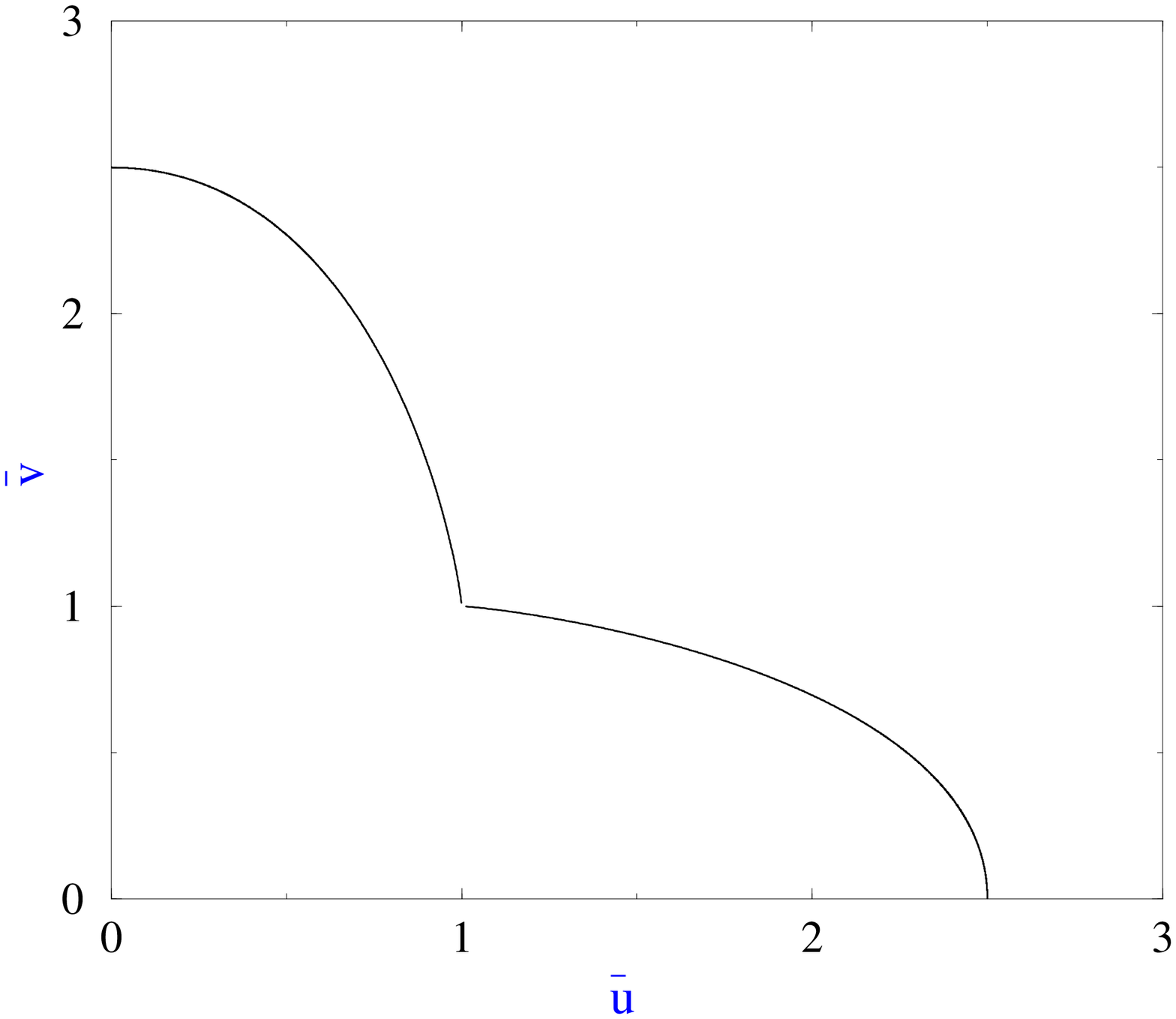}
\caption{\small
Domain of existence of the homogeneous fluid solution ($b=3/2$).
The lower right wing corresponds to $\rho_1=\infty$, $\rho_2$ finite,
and symmetrically for the left upper wing.
}
\label{f7}
\end{center}
\end{figure}

The domain of existence of the homogeneous fluid
solution in the $\u-\v$ plane is shown in figure \ref{f7}.
It is the interior of the domain delimited by the two symmetric curves.
These curves are obtained as follows.
Consider the situation where one of the densities, $\rho_1$, say, is infinite.
Then $v_{k,l}$ is to be taken equal to 1, 
and it is intuitively expected that the two species decouple.
Hence $f_l=(1-\v)\v^l$. It follows that
\be
\rho_2=\frac{\v}{1-\v},
\ee
and
\beq\label{frontiere}
\u=\sum_{l=0}^\infty f_l\left(1+\frac{b}{l}\right)
=1+b(1-\v)(1-\ln(1-\v)).
\eeq
The two former equations give the equation of the boundary of
the domain of existence of fluid solutions, for $\rho_2$ varying.
The second part of the curve is obtained symmetrically by doing the
same analysis with $\rho_2$ infinite.
We note that the tangents to the two curves at the
symmetric point $\bar u=\bar v=1$ are parallel to the axes.

\subsection{Limits of stability of the fluid phase}

Finally the question is how the domain of existence of the fluid region defined 
above is mapped onto the density plane.
As we now show, this domain
maps onto a region of the density plane complementary to a wedge
with tip located at the critical point
$\rho_1=\rho_2=\rho_c$,
as depicted in figure~\ref{f2}. 
All the analysis relies on how the neighbourhood of the point $\u=\v=1$
is mapped onto the density plane.

We begin by a linear analysis of the mapping between the critical points ($\u=\v=1$)
and $(\rho_1=\rho_2=\rho_c)$.
Consider a small segment, with one of the two ends located at $\u=\v=1$,
and with the other one at a given angle with the $\u$ axis.
Let $t$ be the tangent of this angle.
Because of the symmetry between the two species,
and since the transformation of the local derivatives around
the critical point is linear,
the slope of the transformed segment in the density plane is given by
\beq\label{tt}
T=\frac{1+c t}{c+t}.
\eeq
Thus, if $t=\pm1$, then $T=\pm1$.
The constant $c$ is determined numerically by taking $t=\infty$ (segment parallel to the $\v$ axis).
The limiting slopes of the tangents to the wedge at the tip
follow from (\ref{tt}).
For the lower edge it is given by $T(t=0)$, i.e., $1/c$.
For $b=3/2$, we find $T(t=0)\approx0.48$,  with $m_\star=80$.

We then investigate how points 
 located
on the boundary curve~(\ref{frontiere}),  at increasing distances
of the critical point, $\u=\v=1$, are mapped onto the density plane.
For successive values of the cut-off $\max$ the images
of these points are expected to converge to a single curve.
The lower edge of the wedge depicted in  figure~\ref{f2}
is the curve with $m_\star=160$.
The other edge is obtained by symmetry.
These edges represent the limits of stability of the fluid phase.

\section{Condensation and phase diagram}

\subsection{Equal densities}
The case of equal densities, $\rho_1=\rho_2$, is similar to the 
situation encountered for a single-species ZRP~\cite{evans, lux}.
From the analysis of the previous sections, as well as
from numerical integrations of the temporal equations~(\ref{recgene}),
or of their stationary form, the following picture is obtained.

The region $\u=\v<1$ maps onto the fluid phase $\rho_1=\rho_2<\rho_c$, 
corresponding to  exponential solutions of eq.~(\ref{limcont}).
The critical point corresponds to $\u=\v=1$, i.e., $\rho_1=\rho_2=\rho_c$.
Condensation occurs for
$a>3$, i.e. $b>2/3$ (see eq.~(\ref{beq1})), and 
$\rho_1=\rho_2>\rho_c$.
A condensate appears sustaining the excess density
with respect to the critical fluid.
The region $\u=\v>1$ is unphysical.\\

\subsection{Nonequal densities: Existence of a line of transition}

The limits of stability of the condensed phase are given by the 
line $\rho_2=\rho_c$, $\rho_1>\rho_c$, and the symmetric line with respect to
the bisectrix.

There is numerical evidence for the existence of a transition line
between the fluid phase and the condensed phase, lying in between 
the two corresponding stability lines of these phases. 
The transition is discontinuous on this coexistence line.
There is no coexisting solutions accessible dynamically 
on both sides of the line.

The transition between the fluid and condensed phases
is obtained by Monte Carlo simulations of the model,
using the following procedure.
The density $\rho_1$ is fixed to a given value greater than $\rho_c$, and $\rho_2$ increases from a value less than $\rho_c$.
Crossing the stability line of the condensed phase, i.e. for $\rho_2>\rho_c$,
one might expect condensation to occur. 
Instead, the only accessible phase turns out to be the fluid one (see an example of
fluid solution in section~\ref{elim}).
Then, increasing the density $\rho_2$, and crossing the transition line,
there is a sudden phase transition from an imbalanced fluid where both species have densities larger than
the critical density, to a neutral critical fluid and an imbalanced condensate.
Beyond this line, the only accessible solutions are condensed, 
with $\u=\v=1$,
{ while fluid solutions to eq.~(\ref{recgene}) do exist}, as long as $\rho_2$ has
not reached the edge of the wedge.
A surprising consequence of this phase diagram is the occurrence of a
reentrance phenomenon: increasing $\rho_2$ beyond
the symmetric transition line (with respect to the bisectrix) the system becomes fluid again.

\begin{figure}[htb]
\begin{center}
\includegraphics[angle=-90,width=.7\linewidth]{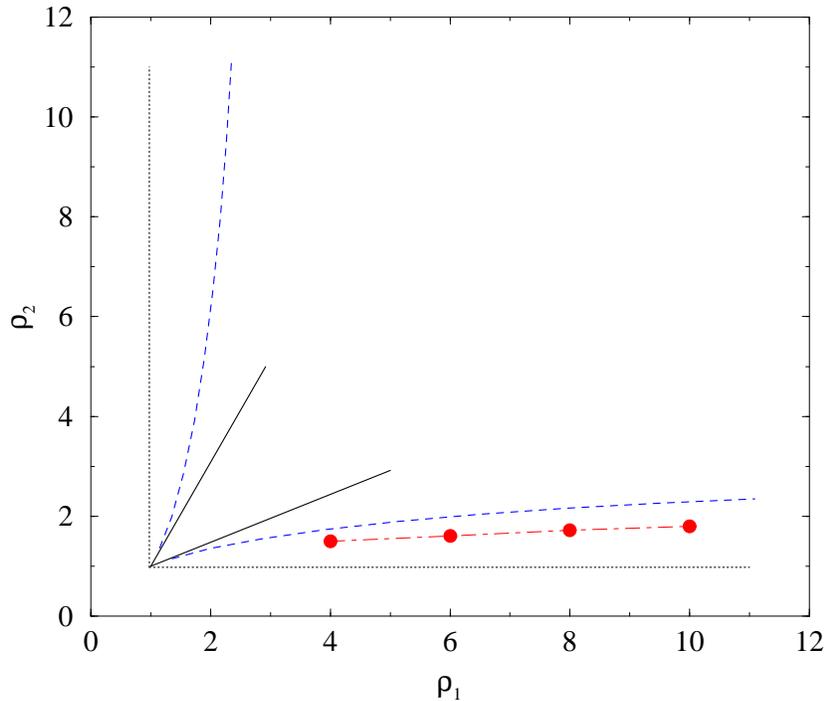}
\caption{\small
Phase diagram in the density plane.
Dot-dashed line with circles: line of transition points
(the symmetric line is not figured).
Dashed lines: limits of stability of the fluid phase.
Dotted lines: limits of stability of the condensed phase.
Straight lines at the tip (critical point) are the local tangents, computed as
explained in the text.
($b=3/2$, $\rho_c\approx0.976$.)}
\label{f2}
\end{center}
\end{figure}
 
We now describe more precisely the method for the determination of the location
of the transition line. 
We fix the value of $\rho_1$,
for example $\rho_1=10$, and let $\rho_2$ increase
from a value less than $\rho_c$.
Then $\u$ and $\v$ are measured. 
Focussing on $\u$,
we observe that, when $\rho_2$ crosses some value, $\rho_2\approx1.8$, there is
a sudden discontinuity in $\u$, dropping from $\u\approx 1.4$ down to
$\u\approx1$.
More precisely, for a system of given size, say $M=40, 60, \ldots$,
the system is first run to stationarity.
Then $n$ successive runs of duration $\Delta t$  less than the flipping time
$\tau$ between the fluid and the condensed phases, and such that $n \Delta t\gg \tau$
are performed.
This flipping time is measured to be exponentially increasing with the system size
(it is approximately doubled
when $M$ is incremented by $20$).
The histogram of the values of $\u$ is a bimodal distribution.
The criterion for the location of the transition point $(\rho_1^*,\rho_2^*)$, when $\rho_2$ varies,
consists in choosing the value of $\rho_2$ such that the two weights of the two maxima are equal.
Thus for $M=40, 60, 80$, we have $\rho_{2}^*\approx1.86, 1.82, 1.8$, 
respectively.

We proceed in the same fashion to obtain the transition points visible
on figure~\ref{f2}, for $\rho_1=4, 6, 8$.
As $\rho_1$ decreases down to $\rho_c$, the discontinuity in $\u$ is smaller
and the determination of the transition point is harder since
it involves larger system sizes.

\section{Final remarks}

Let us first summarize the main outcomes of the present work.
For the process considered, a non-integrable two-species ZRP, we are
able to obtain a number of analytical results from the study of the model
on the complete graph. In particular 
the critical phase is well understood. The coefficient of the exponential decay
of the fluid solutions is analytically predicted.
Finally we can predict the phase diagram of the system by a joint
analytical and numerical investigation.
A salient feature of the phase diagram is the presence
of a single critical point.
A more thorough analysis of the first-order phase transition 
taking place between the fluid phase
and the condensed phase would be interesting, though probably hard to achieve.
The existence of such a transition  
is not expected in integrable
two-species ZRP~\cite{gross,hanney,tom,stefan}.

Beyond the present work, 
a natural question to ask is whether the phenomena observed for the fully
connected geometry survive
in the one-dimensional geometry.
Preliminary investigations
using the analysis performed on the complete graph as a template
indicate similar behaviour.

%
\subsection*{Acknowledgements.}
It is a pleasure to thank J-M Luck for invaluable discussions.
Thanks are also due to EDG Cohen, M Evans, S Grosskinsky, T Hanney,
E Levine and 
D Mukamel for helpful discussions.
This work was partially carried out while CG was a Meyerhoff
Visiting Professor at the Weizmann Institute. Support of the
Albert Einstein Minerva Center for Theoretical Physics and the
Israel Science Foundation (ISF) is gratefully acknowledged.


\section*{References}
\begin{thebibliography}{99}

\bibitem{spitz}
Spitzer F 1970 {\it Advances in Math.} {\bf 5} 246

\bibitem{kip} Kipnis C and Landim C 1999 {\it Scaling limits of interacting particle
systems} Springer

\bibitem{evans} 
Evans M R and Hanney T 2005 \JPA {\bf 38} R195

\bibitem{lux} Godr\`eche C 2006  in {\it Aging and the Glass
Transition} Henkel M, Pleimling M and Sanctuary R eds. 
{\it Springer Lecture Notes in Physics} to be published

\bibitem{wis1}
Kafri Y, Levine E, Mukamel D, Sch\"utz G M and T\"{o}r\"{o}k J 2002
\PRL \textbf{89} 035702

\bibitem{glm}
Godr\`eche C, Levine E and Mukamel D 2005 \JPA {\bf 38} L523

\bibitem{andj}
Andjel E D 1982 {\it Ann. Prob.} {\bf 10} 525

\bibitem{urn} For a short review, see 
Godr\`eche C and Luck J-M 2002 {\it J. Phys. Cond. Matt.} {\bf 14} 1601

\bibitem{gross} Grosskinsky S and Spohn H 2003
{\it Bull. Braz. Math. Soc.} {\bf 34}  489

\bibitem{hanney} Evans M R and Hanney T 2003 \JPA {\bf 36} L44

\bibitem{kol}Kolmogorov A N 1936 {\it Math Ann.} {\bf 112} 115

\bibitem{kel}Kelly F 1979 {\it Reversibility and Stochastic Networks} Wiley

\bibitem{tom}
Hanney  T and Evans M R 2004 {\it Phys. Rev. E} {\bf 69} 016107

\bibitem{stefan} Grosskinsky S in preparation

\end{thebibliography}
\end{document}